\documentclass[a4paper]{jpconf}
\usepackage{graphicx}
\usepackage{amsmath}
\usepackage{hyperref}

\begin{document}
\title{Reweighting with Boosted Decision Trees}

\author{Alex Rogozhnikov$^{1,2}$}

\address{$^1$ National Research University Higher School of Economics (HSE), RU}
\address{$^2$ Yandex School of Data Analysis (YSDA), RU}

\ead{axelr@yandex-team.ru}

\begin{abstract}
Machine learning tools are commonly used in modern high energy physics (HEP) experiments.
Different models, such as boosted decision trees (BDT) and artificial neural networks (ANN), are widely used in analyses and even in the software triggers~\cite{trigger}. 

In most cases, these are classification models used to select the ``signal'' events from data.
Monte Carlo simulated events typically take part in training of these models.
While the results of the simulation are expected to be close to real data, 
in practical cases there is notable disagreement between simulated and observed data.
In order to use available simulation in training, corrections must be introduced to generated data.
One common approach is reweighting --- assigning weights to the simulated events.

We present a novel method of event reweighting based on boosted decision trees.
The problem of checking the quality of reweighting step in analyses is also discussed.

\end{abstract}

\section{Introduction}

Reweighting distributions (also known as event reweighting or importance weighting) is a general procedure,
but its major use-case for particle physics is to modify the output of the Monte Carlo (MC) simulation
to reduce disagreement with real data (RD) collected at a collider.

There are many applications in HEP including searching for rare decays (decays with an extremely low probability in the standard model of elementary particles), 
when a classifier is trained on MC data to discriminate signal decays from background.
However, the simulation is often imperfect (see~\cite{MVAreweighting} for more details) and corrections should be introduced.
To calibrate the reweighting a similar physics process is considered, for which both real data and simulation can be obtained.
For instance, in rare decays a normalization channel is selected --- a decay
with the similar kinematic characteristics (see~\cite{kaggleChallenge} for an example)\footnote{
    Data-MC inconsistencies can also be taken into account 
    by the calibration of subdetector response, 
    Monte Carlo generator tunes \cite{tunes} 
    or discarding regions, where disagreement is the worst.
}.

Reweighting techniques have applications outside HEP: i.e.\ in sociology a survey reweighting is used to reduce a non-response bias~\cite{sociologyReweighting}.  
In what follows HEP terminology is used, but approaches discussed in the paper are applicable to any reweighting.

Mathematically, the problem is equivalent to estimating the density ratio
${f_\text{RD}(x)} / {f_\text{MC}(x)}$ as a function of the variables participating in reweighting.
A density ratio estimation is a general problem in machine learning (ML) with numerous applications (see \cite{densityRatio}).

\section{Basic approach to event reweighting}

An approach widely used in High Energy Physics is reweighting with bins. 
The space of variables is split into bins, in each bin the weights of the simulated events are multiplied by 
\[
	\text{multiplier}_\text{bin} = \dfrac{w_\text{bin, RD}}{w_\text{bin, MC}}
\]
to compensate the difference ($w_\text{bin, RD}$ and $w_\text{bin, MC}$ ---  total weight of events in a bin for RD and MC distributions).
In other words, both densities $f_\text{RD}(x)$ and $f_\text{MC}(x)$ are estimated using histograms and then divided (this gives another name of this approach --- ``histogram division'').

Reweighting using bins is intuitive and easy-to-use, however, has very strong limitations:
\begin{itemize}
	\item 
	very few variables can be reweighted in practice, typically one or two;
	\item 
	choosing which variable(s) to use in reweighting is complex: reweighting one variable often brings disagreement in others;
	\item 
	the amount of data needed to reliably estimate a density function with a histogram grows exponentially with the number of variables, which is commonly referred to as the ``curse of dimensionality''.
\end{itemize}
To fight the last problem, one can reduce the number of bins along each variable, but this drives to a rough reweighting rule, insufficient to cover discrepancies.

\section{Reusing classification ML techniques to reweight distribution}

Density estimation is a complex problem and it should be avoided in cases when only the ratio is of interest.
A general and natural method of density ratio estimation is based on reusing general-purpose ML techniques.
In~\cite{MVAreweighting} this method was proposed and was successfully applied to particle physics problems.

Some general-purpose classification techniques (i.e.\ BDT and ANN) trained to discriminate MC and RD can provide probabilities $p_\text{MC}(x)$, $p_\text{RD}(x)$ that a given event $x$ belongs to MC or RD. 
The probabilities can be used to estimate required density ratio:
\[
	\dfrac{f_\text{RD}(x)}{f_\text{MC}(x)}
	\sim 
	\dfrac{p_\text{RD}(x)}{p_\text{MC}(x)}.
\]

This approach successfully overcomes the curse of dimensionality, but provides inaccurate predictions when density ratio is high. 
One possible explanation is as follows:
while regions with the high ratio are significant for reweighting,
those are not of high importance for classification task:
guessing correct class within regions with high / low ratio is easier, since most of the events belong to one class,
and classification algorithms focus on the other regions.
For example, when training ANN or GBDT, these regions provide smaller contribution to the loss function,
thus are given less attention.

\section{BDT reweighter}

In this section a machine learning algorithm is proposed to solve the specific problem of reweighting. To address the problems of the histogram reweighting approach, the space of variables is split into a few large regions.
These regions are not obtained by a simple splitting of each variable into several bins, but in correspondence with the problem.

A decision tree is used to split the regions.
Recall that decision trees split the space of variables into the regions by checking simple conditions. 
Each region is associated with some leaf of the tree.

To find the regions that are suitable for reweighting, the symmetrized~$\chi^2$ is
greedily optimized\footnote{For comparison:
    in the gradient boosting algorithm decision trees are built by greedily minimizing mean squared error.
}:
\begin{equation}
\chi^2 = \sum_\text{leaf}
	\dfrac{(w_\text{leaf, MC} - w_\text{leaf, RD})^2}{w_\text{leaf, MC} + w_\text{leaf, RD}}.
\end{equation}
This metric is maximized to find the regions important for reweighting.
If in some leaf (region) the amount of MC events $w_\text{leaf, MC}$ is much higher than the amount $w_\text{leaf, RD}$ of RD events, the MC weights in this region must be decreased. 
The corresponding summand in the $\chi^2$ will be high reflecting the importance of this region for reweighting.

The BDT reweighter makes use of many such trees which are trained one-by-one by repeating the following steps many times:
\begin{enumerate}
	\item
	build a shallow tree to maximize the symmetrized $\chi^2$
	\item
	compute predictions in the leaves:
	$ \text{leaf\_pred} = \log \dfrac{w_\text{leaf, MC}}{w_\text{leaf, RD}} $
	\item
	reweight the MC distribution (compare this step with AdaBoost~\cite{adaboost}): 
	\begin{equation}
	w = \begin{cases}
		w, & \text{if event from RD distribution} \\
		w \times e^\text{pred}, & \text{if event from MC distribution}
		\end{cases}
	\end{equation}
	For each event $\text{pred}$ is equal to the prediction of a leaf containing this event.
\end{enumerate}

The last two steps work in the same way as reweighting with bins, the distinction being that the bins are selected differently.
Also, since logarithm is taken, the predictions of the different trees are summed up, as is usually done in the boosting.

In the BDT reweighter, each tree in the sequence is trying to cover the discrepancies that were not resolved on the previous iterations. 
The complexity of a decision tree can be adjusted by varying the depth and the minimal number of samples in the leaf, making this approach highly tunable.

\section{Comparison of multidimensional distributions}


The goal of reweighting is to have the MC distribution coincide with the RD distribution.

Comparing one-dimensional distributions is simple and can be done either by looking at the distributions or by computing one of well-known two-sample tests like Kolmogorov-Smirnov, Anderson-Darling or Cramer--von Mises. 
However, in the applications all the distributions are multidimensional. 
Comparing only projections is obviously not enough to be sure that the distributions are identical.

At the same time, there are no useful multidimensional two-sample tests. Given the whole pipeline of our analysis, two-samples tests are not necessary, because the question of interest is not whether the MC and RD distributions are different (those are different).
The question is whether an ML technique used later in the analysis (i.e.\ to detect signal decay, typically it is BDT or ANN) is able to use the discrepancy between RD and MC.
Thus it is only needed to check that after reweighting a classifier used in the analysis is not able to find the difference between the distributions. 
For this purpose a classifier is trained to discriminate RD and MC.\footnote{This
    also gives another perspective of reweighting as an adversarial process, involving discriminator and reweighter.
	Training of adversarial neural networks is actively studied now, see \cite{adversarial}.
}
Its quality is checked by
inspecting the ROC curves on a holdout sample.\footnote{See
    also~\cite{friedman} for an alternative ML-based approach to compare multidimensional distributions.
}

Overfitting is an issue that becomes obvious when using advanced methods of reweighting.
One should measure the quality of the classification on a holdout (data sample that was not participating in training) to get unbiased estimations. 
The same approach works for reweighting: the quality of reweighting should be checked on the data that did not participate in training of the reweighting rule. 
The different cross-validation techniques like folding are also applicable.

\section{Case study}

\begin{figure}
	\begin{center}
		\begin{minipage}{17pc}
			\includegraphics[width=17pc]{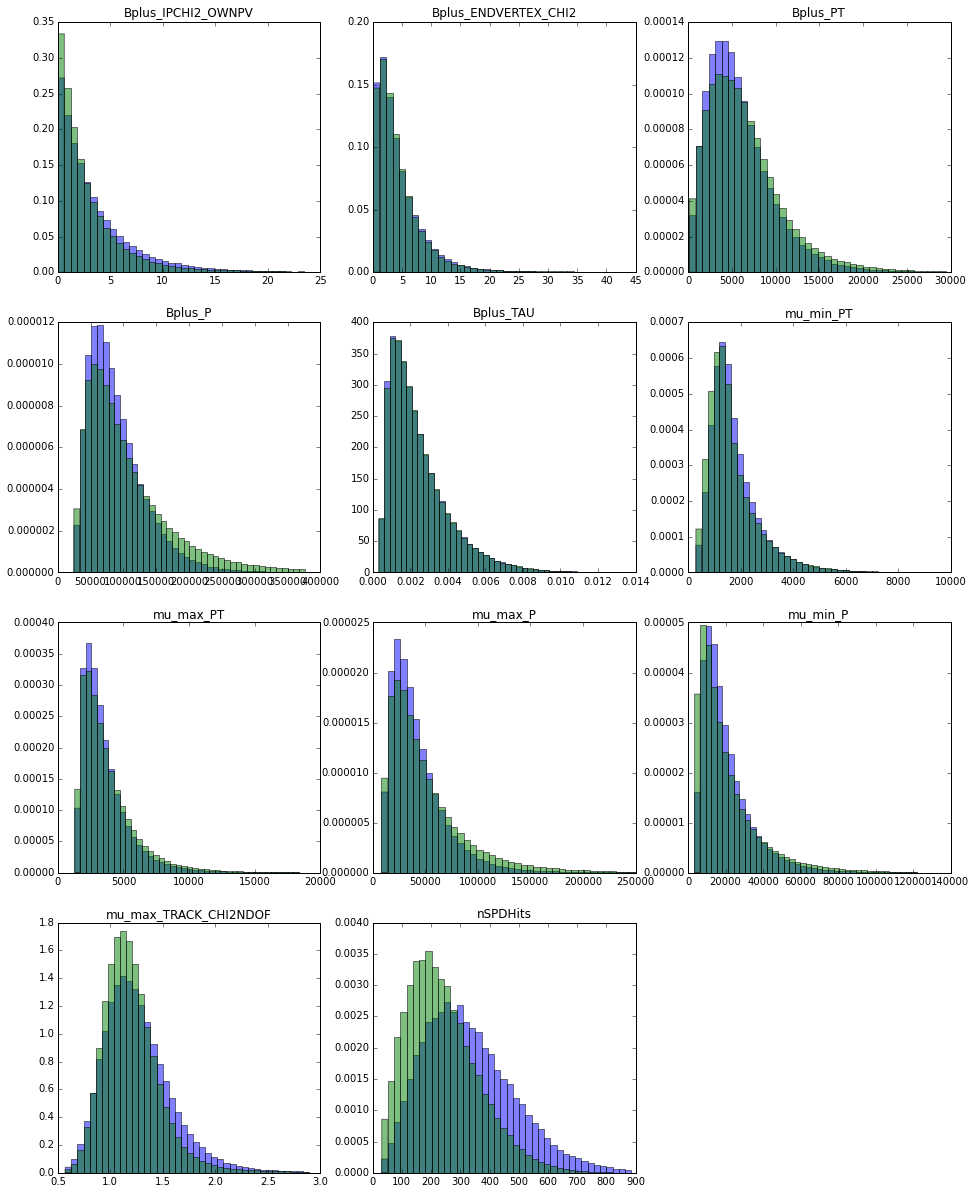}
		\end{minipage}
		\hspace{1cm}
		\begin{minipage}{17pc}
			\includegraphics[width=17pc]{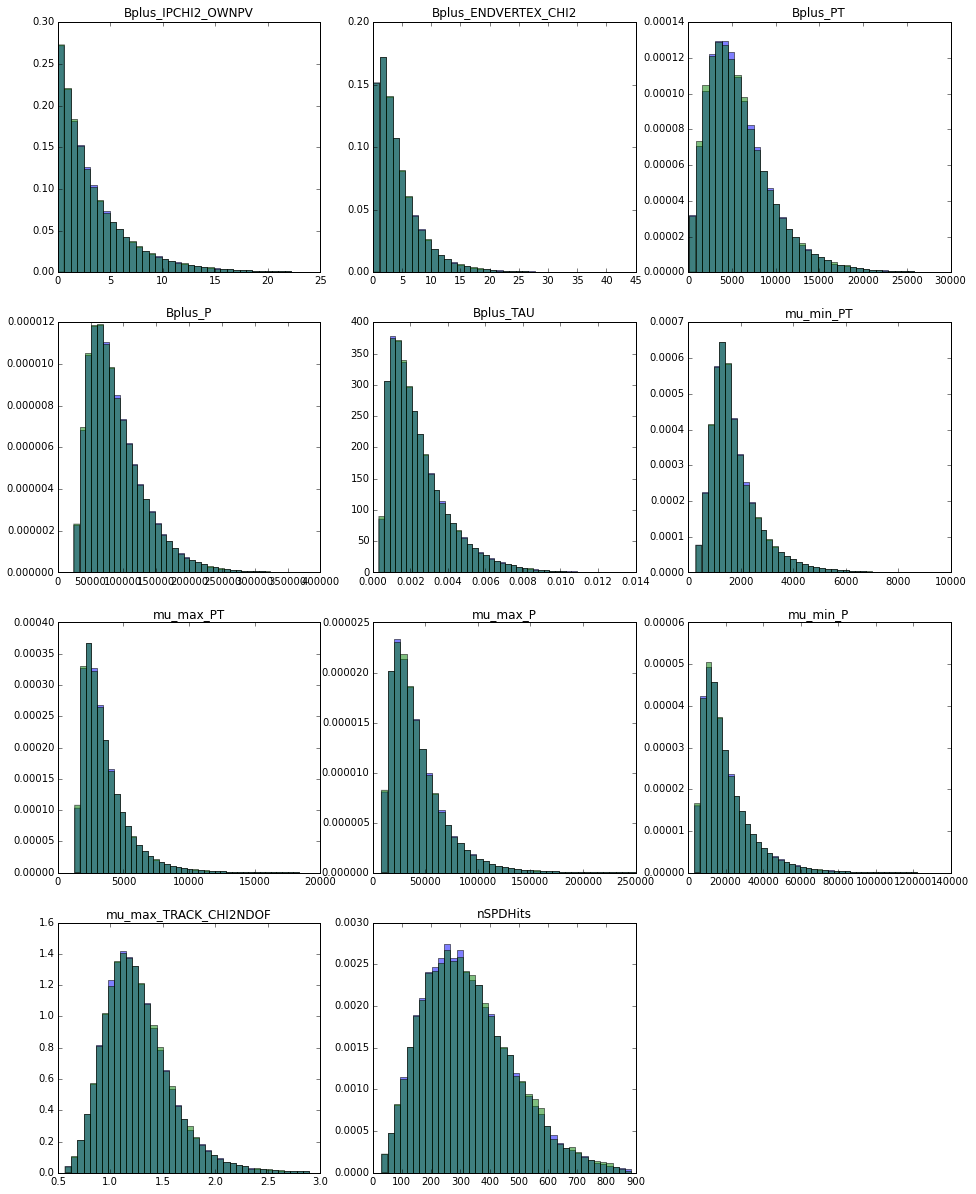}
		\end{minipage} 
		\caption{\label{fig:reweightDistrib} 
		Comparison of real data (blue) and simulated (green) distributions before and after using the BDT reweighter.
		}
		\vspace{-0.3cm}
	\end{center} 
\end{figure}

As an example the 11-dimensional distribution is taken (simulated and real data). 
Figure~\ref{fig:reweightDistrib} demonstrates how the distribution of different features (variables) has changed after reweighting with the new method.

\begin{figure}

	\begin{tabular}{l|r|r|r|r}
	\hline
	Feature name &  original &  bins reweight &  reuse ML &  BDT reweighter \\
	\hline
	Bplus\_IPCHI2\_OWNPV	& 0.0796	& 0.0642	& 0.0463	& 0.0028 \\
	Bplus\_ENDVERTEX\_CHI2	& 0.0094	& 0.0175	& 0.0490	& 0.0021 \\
	Bplus\_PT				& 0.0586	& 0.0679	& 0.0126	& 0.0053 \\
	Bplus\_P				& 0.1093	& 0.1126	& 0.0044	& 0.0047 \\
	Bplus\_TAU				& 0.0037	& 0.0060	& 0.0324	& 0.0044 \\
	mu\_min\_PT				& 0.0623	& 0.0604	& 0.0017	& 0.0036 \\
	mu\_max\_PT				& 0.0483	& 0.0561	& 0.0053	& 0.0035 \\
	mu\_max\_P				& 0.0906	& 0.0941	& 0.0084	& 0.0036 \\
	mu\_min\_P				& 0.0845	& 0.0858	& 0.0058	& 0.0043 \\
	mu\_max\_TRACK\_CHI2NDOF	& 0.0956	& 0.0042	& 0.0128	& 0.0043 \\
	nSPDHits				& 0.2478	& 0.0098	& 0.0180	& 0.0075 \\
	\hline
	\end{tabular}
	\caption{
	\label{fig:KSdistances}
	Kolmogorov-Smirnov distances for each variable before reweighting and after applying different reweighting techniques. 
	Only the last two variables are used during reweighting with bins.
	}
\end{figure}

\begin{figure}
	\begin{center}
		\begin{minipage}{0.4\textwidth}
			\includegraphics[width=15pc]{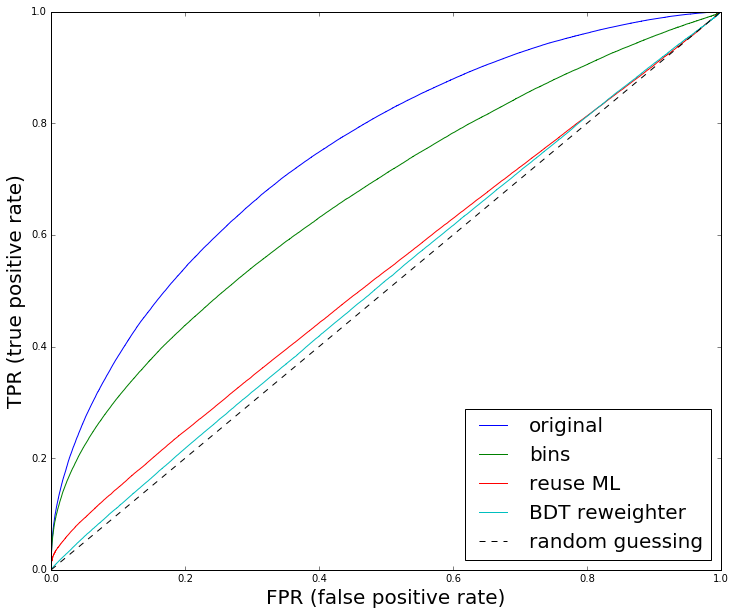}
		\end{minipage}
		\begin{minipage}{0.5\textwidth}
		\caption{\label{fig:rocCurves} 
		Checking the quality of reweighting with ML. 
		The ROC curves for the classifier trained to discriminate RD and MC are computed on a holdout. 
		Reweighting with bins significantly reduces initially high discrepancy, 
		but the classifier still can easily find the difference. 
		ML-based solutions provide significantly better results, though reusing ML approach has minor issues in the left bottom corner.
		}
		\end{minipage}
	\end{center} 
\end{figure}

In table \ref{fig:KSdistances} the Kolmogorov-Smirnov distances are provided. 
Reweighting with bins is done for last two variables, while two other approaches use all 11 variables.
Finally, the quality of reweighting is checked as it was proposed earlier and the ROC curves are built on a holdout (figure \ref{fig:rocCurves}).

\section{Conclusion}

Two problems are discussed in the paper:
\begin{enumerate}
	\item event reweighting for multidimensional distributions
	\item the comparison of multidimensional distributions
\end{enumerate}
It is demonstrated that both problems are effectively addressed by means of machine learning, while typically these steps in the analysis are considered outside of the scope of ML.

Also, the novel method of reweighting is proposed: a modification of BDT algorithm, which alters the procedures of boosting and decision tree building. 
This method outperforms known reweighting approaches and makes it possible to reweight dozen of variables. When compared on the same problems, it requires less data to achieve the same quality.

Ready-to-use implementation of introduced algorithm is available in the \verb|hep_ml| package \cite{githubHepML}.


\section*{References}
\begin{thebibliography}{99}
	\bibitem{trigger}
	Likhomanenko, T., Ilten, P., Khairullin, E., Rogozhnikov, A., Ustyuzhanin, A., Williams, M. (2015). LHCb Topological Trigger Reoptimization. In Journal of Physics: Conference Series (Vol. 664, No. 8, p. 082025). IOP Publishing.

	\bibitem{MVAreweighting}
	Martschei, D., Feindt, M., Honc, S.,  Wagner-Kuhr, J. (2012). Advanced event reweighting using multivariate analysis. In Journal of Physics: Conference Series (Vol. 368, No. 1, p. 012028). IOP Publishing.

	\bibitem{kaggleChallenge}
	Blake, T., Bettler, M.-O., Chrzaszcz, M., Dettori, F., Ustyuzhanin, A., and Likhomanenko, T. (2015). 
	Flavours of physics: the machine learning challenge or the search of $\tau \to 3\mu$ decays at LHCb

	\bibitem{tunes}
	ATLAS collaboration. (2012). Summary of ATLAS Pythia 8 tunes (Vol. 14). ATL-PHYS-PUB-2012-003.

	\bibitem{sociologyReweighting}
	Kizilcec, R. Reducing non-response bias with survey reweighting: Applications for online learning researchers. Proceedings of the first ACM conference on Learning @ scale conference. ACM, 2014.

	\bibitem{densityRatio}
	Sugiyama, M., Suzuki, T., Kanamori, T. 
	Density Ratio Estimation in Machine Learning, 
	Cambridge University Press, Cambridge, UK, 2012. 

	\bibitem{friedman}
	Friedman, J. (2003). On Multivariate Goodness-of-Fit and Two-Sample Testing. In Statistical Problems in Particle Physics, Astrophysics, and Cosmology (Vol. 1, p. 311).

	\bibitem{adversarial}
	Goodfellow, I., Pouget-Abadie, J., Mirza, M., Xu, B., Warde-Farley, D., Ozair, S., Courville, A., Bengio, Y. (2014). Generative adversarial nets. Advances in Neural Information Processing Systems (pp. 2672-2680).

	\bibitem{adaboost}
	Freund, Y., Schapire, R. E. (1996). Experiments with a new boosting algorithm. In ICML (Vol. 96, pp. 148-156).

	\bibitem{githubHepML} 
	\url{https://github.com/arogozhnikov/hep_ml}

\end{thebibliography}
\end{document}